\documentclass[fleqn,twoside]{article}
\usepackage{espcrc2}

\usepackage{graphicx}
\usepackage[figuresright]{rotating}

\newcommand{\AmS}{{\protect\the\textfont2
  A\kern-.1667em\lower.5ex\hbox{M}\kern-.125emS}}

\title{Massive neutrinos and dark energy}

\author{Paolo Serra,\address{Physics Department, University of Rome ``La
 Sapienza'' and INFN - Sezione di Roma}
 Rachel Bean\address{Dept. of Astronomy, Cornell University, Ithaca, NY 14853}, Axel De La Macorra\address{Instituto de F\'{\i}sica, UNAM, Apdo. Postal 20-364,
01000 M\'exico D.F., M\'exico}, Alessandro Melchiorri$^{a}$}
\begin{document}

\begin{abstract}
We consider the impact of the Heidelberg-Moscow claim for a detection
of neutrino mass on the determination of the dark energy equation of state.
By combining the Heidelberg-Moscow result with the WMAP 3-years
data and other cosmological datasets we constrain the equation of
state to $-1.67< w <-1.05$ at $95 \%$ c.l., 
While future data are certainly needed for a confirmation of the
controversial Heildelberg-Moscow claim, our result shows that
future laboratory searches for neutrino masses may play a crucial
role in the determination of the dark energy properties.
\vspace{1pc}
\end{abstract}

\maketitle
\section{Introduction}
As very well known in the literature massive neutrinos
can be extremely relevant for cosmology and leave key signatures
in several cosmological data sets. 
 Current cosmological data, in the framework of
a cosmological constant, are able to indirectly constrain the
absolute neutrino mass to $\Sigma m_{\nu} < 0.75\,\mathrm{eV}$ at $95 \%$ c.l.
\cite{wmap3cosm} and are in tension with the Heidelberg-Moscow claim
(HM hereafter). In fact, double beta decay searches from the HM experiment
have reported a signal for a neutrino mass
at $>4\sigma$ level \cite{Kl04}, recently
promoted to $>6\sigma$ level by a pulse-shape analysis \cite{Kl06}.
This claim translates in a
total neutrino mass of $\Sigma m_{\nu} > 1.2\,\mathrm{eV}$ at $95 \%$ c.l..
While this claim is still considered as controversial (see e.g. \cite{Elli}),
it should be noted that it comes from the most sensitive ($^{76}$Ge)
detector to date and no independent experiment can, at the moment,
falsify it.

However, as first noticed by \cite{hannestad}, there is some
form of anticorrelation between the equation of state parameter
$w$ and $\Sigma m_{\nu}$.
The cosmological bound on neutrino masses can therefore be relaxed
by using a DE component with a more negative
value of $w$ than a cosmological constant. As we show here, the HM claim
is compatible with the cosmological data only if
the equation of state (parameterized as constant) is
$w < -1$ at $95 \%$.

\section{Method}
The method we adopt is based on the publicly available Markov Chain Monte Carlo
package \texttt{cosmomc} \cite{Lewis:2002ah} with a convergence
diagnostics done through the Gelman and Rubin statistic.
We sample the following eight-dimensional set of cosmological
parameters,
adopting flat priors on them: the physical baryon, Cold Dark Matter
and massive neutrinos densities,
$\omega_b=\Omega_bh^2$,
$\omega_c=\Omega_ch^2$ and $\Omega_{\nu}h^2$,
 the ratio of the sound horizon to the angular diameter
distance at decoupling, $\theta_s$, the scalar spectral index $n_s$,
the overall normalization of the spectrum $A$ at
$k=0.05$ Mpc$^{-1}$, the optical
depth to reionization, $\tau$, and, finally, the
DE equation of state parameter $w$.
Furthermore, we consider purely adiabatic initial conditions and we
impose flatness.

We include the three-year WMAP data \cite{wmap3cosm} (temperature
and polarization) with the routine for computing the likelihood
supplied by the WMAP team. Together with the WMAP data we also
consider the small-scale CMB measurements of CBI
\cite{2004ApJ...609..498R}, VSA \cite{2004MNRAS.353..732D}, ACBAR
\cite{2002AAS...20114004K} and BOOMERANG-2k2
\cite{2005astro.ph..7503M}.  In addition to the CMB data, we include
the constraints on the real-space power spectrum of galaxies from
the SLOAN galaxy redshift survey (SDSS) \cite{2004ApJ...606..702T}
and 2dF \cite{2005MNRAS.362..505C},
 and the Supernovae Legacy Survey data from \cite{2006A&A...447...31A}.
Finally, we include the Heidelberg-Moscow as in the recent analysis of
\cite{fogli2}. Fore a more detailed description of the analysis see \cite{macorra}.

Using the theoretical input for
$C_{mm}({}^{76}\,\mathrm{Ge})$ from Ref.~\cite{faessler}, the
$0\nu2\beta$ claim of \cite{Kl04} is transformed in the
%following
$2\sigma$ range
\begin{equation}
\log_{10}(m_{\beta\beta}/\mathrm{eV})= -0.23\pm 0.14
\label{logmbb2}\ ,
\end{equation}
i.e., $0.43<m_{\beta\beta}<0.81$ (at $2\sigma$, in $\mathrm{eV}$).

Considering all current oscillation data (see \cite{fogli2}) and
under the assumption of a $3$ flavor neutrino mixing the above
constraint yields:
\begin{equation}
0.0137< \Omega_{\nu}h^2 <0.026
\end{equation}
at $95 \%$ c.l. where we used the well known relation:
${\Omega_{\nu}h^2}=\Sigma m_{\nu}/93.2\,\mathrm{eV}$.

Our main results are plotted in Fig.1 where we show the constraints on
the $w-\Sigma m_{\nu}$ plane in two cases,
 with and without the HM prior on
neutrino masses. As we can see, without the
HM prior we are able to reproduce the results already
presented in the literature (see e.g. \cite{wmap3cosm}),
namely current cosmological data constrain  neutrino masses
to be $\Sigma m_{\nu} <0.75\,\mathrm{eV}$.
However an interesting anti-correlation is present between the DE
parameter $w$ and the neutrino masses and larger
neutrino masses are in better agreement with the data
for more negative values of $w$.
It is therefore clear that when we add the HM prior
($\Sigma m_{\nu} \sim 1.8 \pm 0.6\,\mathrm{eV}$ 
at $95 \%$ c.l., again see Fig.1)
the contours are shifted towards higher
values of neutrino masses and towards lower values
of $w$. A combined analysis of cosmological
data with the HM priors gives
$-1.67< w <-1.05$
and $0.66<\Sigma m_{\nu}<1.11$ (in $\mathrm{eV}$)
at $95 \%$ c.l. excluding the case of the cosmological constant
at more than $2 \sigma$ with
$\Sigma m_{\nu} = 0.85\,\mathrm{eV}$, $w =-1.31  $ and $\Omega_m=0.35$
as best fit values.
Without the HM prior the data gives $-1.28 < w < -0.92$
and $\Sigma m_{\nu} < 0.73\,\mathrm{eV}$
again at $95 \%$ c.l. with $w=-1.02$, $\Sigma m_{\nu} = 0.05\,\mathrm{eV}$
and $\Omega_m=0.29$  as best fit
 values.

The inclusion of the HM prior affects also
other parameters. We found, at $95 \%$ c.l.:
  $0.916 < n_s < 0.979$
($0.926 < n_s < 0.989$ withouth HM), $0.0209 < \Omega_bh^2 < 0.0235$
($0.0211 < \Omega_bh^2 < 0.0238$ without HM),
$0.302 < \Omega_m < 0.444$ ($0.262 < \Omega_m < 0.360$ without HM).
It is interesting to notice that the inclusion of massive
neutrinos seems to further rule out the scale-invariant $n_{s}=1$
model.
\begin{figure}[tbp]
    \includegraphics[width=8.0cm]{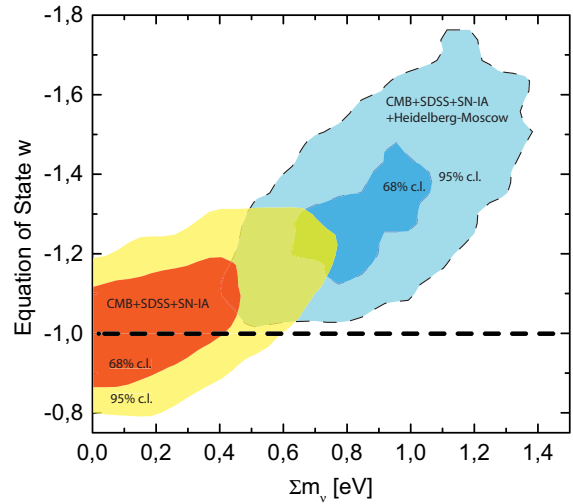}
  \caption{Constraints on the $w-\Sigma$ plane in two cases
with and without the Heidelberg-Moscow prior on
neutrino masses. Reprinted from \cite{macorra}.}
 \label{fig1}
\end{figure}
From a theoretical point of view, there are several possibilities for
an equation of state of dark energy less than $-1$.
Scalar fields with positive kinetic energy have $w>-1$
while phantom fields \cite{phantom} can have $w<-1$
but they have a negative kinetic energy and
many fundamental theoretical problems.
Another possiblity is to follow the approach of interacting
 DE \cite{IDE}-\cite{macorra}. Interacting DE are models where the dark
 energy interacts with other particles, as for example dark
 matter or neutrinos. In fact, the energy
scale of DE (${\cal  O}(10^{-3})$~$\mathrm{eV}$) is of the order of the
neutrino mass scale and this may suggest for a link between neutrino
physics and DE that must certainly be further investigated. In particular,
 the net effect of this interaction is to change  the apparent
 equation of state of DE\cite{IDE}. 
%An observer that supposes
% that the DE has no interaction sees a different
% evolution of DE as an observer that takes into account
%for the interaction of DE. This effect allows to have an  apparent
%equation of state $w<-1$ for the ``non-interaction" DE observer even though
%the true equation of state of the DE is larger than -1.
\section{Conclusions}
We have considered the cosmological implications of
the controversial HMresult. A scenario based on a cosmological
constant is unable to provide a good fit to current data when
a massive neutrino component as large as suggested by HM is included
in the analysis. A better fit to the data is obtained when
the DE component is described with an equation of state
$w \sim -1.3$, with $w < -1$ at more than $95 \%$ c.l..
As far as we know, this is the only dataset able to exclude
a cosmological constant at such high significance.

There exists, therefore, a significant tension between the indirect,
observational measurements leading to the LCDM scenario and the direct
HM observations. Rather than implying one should rule out evidence
from the direct measurements purely on the basis of disparity with
the indirect observations, this tension suggests we should keep
our minds open to alternative dark energy scenarios beyond a
cosmological constant.

Systematics can be present in the HM data and a more conservative
treatment (see \cite{Kl04}) would lead
to a better agreement with a cosmological constant.
However, phantom models with $w < -1$ would still provide
a better fit to the data.
On the other hand, using a more conservative
approach towards cosmology, by, for example, combining HM only with
the CMB dataset, would provide even larger values of $\Sigma m_{\nu}$
and more negative values for $w$.
Recent combined analysis with Lyman-$\alpha$ forest data
(\cite{seljakanze},\cite{fogli2})
imply tight constraints on neutrino masses ($\Sigma m_{\nu} <0.2\,\mathrm{eV}$),
seemingly at discord with the HM result, and also in some tension with
CMB data alone. Future larger scale Lyman-$\alpha$
surveys and refinements in the analysis, addressing systematic
uncertainties and sensitivity to modeling assumptions,
will allow a better assessment of how these tensions will be resolved.

A determination of the absolute neutrino mass scale will therefore
not only bring relevant information for neutrino physics but may be
extremely important in the determination of the dark energy
properties and in shedding light on a possible neutrino-dark energy
connection. 
%Future direct particle detection and indirect astronomical
% experiments will scrutinize this interesting hypothesis.

\noindent {\bf ACKNOWLEDGMENTS}\\
It is a pleasure to thank the organizers of the NOW 2006 conference
for a beautiful and stimulating workshop. 
%I also thank my
%collaborators who worked with me on this project discussed in more
%detail in \cite{macorra}


\begin{thebibliography}{9}
%\bibitem{pastor} J. Lesgourgues and S. Pastor,
%                Phys. Rept. {\bf 429}, 307 (2006)
\bibitem{wmap3cosm} D. N. Spergel {\it et al.}, Arxiv:
  astro-ph/0603449 (2006).

\bibitem{Kl04}  H. V.~Klapdor-Kleingrothaus {\em et al.}
                Phys.\ Lett.\ B 586 (2004) 198.

\bibitem{Kl06}  H. V.~Klapdor-Kleingrothaus, talk at {\em SNOW 2006},
                2nd Scandinavian Neutrino Workshop (Stockholm, Sweden,
                (2006)).

\bibitem{Elli}  S.R.~Elliott and J.~Engel,
                J.\ Phys.\ G 30 (2004) R183.

\bibitem{hannestad}  S.~Hannestad,
 %cosmological neutrino mass bound,''
  Phys.\ Rev.\ Lett.\ 95 (2005) 221301.
  %%CITATION = ASTRO-PH 0505551;%%

\bibitem{seljakanze}
 U.~Seljak, A.~Slosar and P.~McDonald,
  %galaxy clustering and SN constraints,''
  Arxiv: astro-ph/0604335 (2006).

\bibitem{IDE}
  W.~Zimdahl and D.~Pavon, Phys.\ Lett.\ B 521 (2001), 133;
  D.~B.~Kaplan {\em et al.} Phys.\ Rev.\ Lett.\  93 (2004) 091801;
  R.~D.~Peccei,
  Phys.\ Rev.\ D 71 (2005) 023527;
  S.~Das, P.~S.~Corasaniti and J.~Khoury,Phys.\ Rev.\ D 73 (2006) 083509;
 A.~W.~Brookfield {\em et al.} Phys.Rev.D 73 (2006) 083515;
 M.~Kaplinghat and A.~Rajaraman, arXiv:astro-ph/0601517 (2006).

\bibitem{Lewis:2002ah}
A.Lewis and S.Bridle,
Phys.\ Rev.\ D 66 (2002) 103511.

\bibitem{phantom}
  S.~M.~Carroll, M.~Hoffman and M.~Trodden,
Phys.\ Rev.\ D 68 (2003) 023509.

\bibitem{2004ApJ...609..498R}
A.~C.~S.\ Readhead {\em et al.}, ApJ  609 (2004) 498. 

\bibitem{2004MNRAS.353..732D}
C.\ Dickinson {\em et al.}, MNRAS 353 (2004) 732.

\bibitem{2002AAS...20114004K}
C.~L.\ Kuo {\em et al.}, American Astronomical Society
  Meeting, Vol. (2002) 201.

\bibitem{2005astro.ph..7503M}
C.~J.\ MacTavish {\em et al.}, arXiv:astro-ph/0507503 (2005).

\bibitem{2004ApJ...606..702T}
M.\ Tegmark {\em et al.}, ApJ 606 (2004) 702.

\bibitem{2005MNRAS.362..505C}
S.\ Cole {\em et al.}, MNRAS 362 (2005) 505.

\bibitem{2006A&A...447...31A}
P.\ Astier {\em et al.}, A\&A 447 (2006) 31.

\bibitem{fogli2}  G.~L.~Fogli {\it et al.},
  %WMAP-3y and first MINOS results,''
   arXiv:hep-ph/0608060 (2006).
  %%CITATION = HEP-PH 0608060;%%
\bibitem{macorra} A.~De La Macorra {\em et al.} 
- Arxiv preprint astro-ph/0608351 (2006),
in press.

\bibitem{faessler}  V.~A.~Rodin {\em et al.}, Nucl.\ Phys.\ A 766
  (2006) 107.



\end{thebibliography}
\end{document}